\documentclass[referee]{aa}
\usepackage[varg]{txfonts}
\usepackage{graphicx}
\usepackage{natbib}
\bibpunct{(}{)}{;}{a}{}{,} 

\begin{document}

\title{The AD775 cosmic event revisited: the Sun is to blame}

\author{I.G. Usoskin\inst{1}
\and B. Kromer\inst{2}
\and F. Ludlow\inst{3}
\and J. Beer\inst{4}
\and M. Friedrich\inst{5}
\and G. A. Kovaltsov\inst{6}
\and S. K. Solanki\inst{7,8}
\and L. Wacker\inst{9}}

\institute{Sodankyl\"a Geophysical Observatory (Oulu unit) and Physics Dept., University of Oulu, Finland
\and Klaus-Tschira-Laboratory for Scientific Dating, Curt-Engelhorn-Centre for Archaeometry, D6, 3, 68159 Mannheim, Germany
\and Harvard University Center for the Environment, and Department of History, Harvard University, Cambridge, Massachusetts, USA
\and Swiss Federal Institute of Aquatic Science and Technology, Eawag, \"Uberlandstrasse 133, 8600 D\"ubendorf, Switzerland
\and Hohenheim University, Institute of Botany (210), D-70593 Stuttgart, Germany
\and Ioffe Physical-Technical Institute, 194021 St. Petersburg, Russia
\and Max Planck Institute for Solar System Research, Max-Planck-Str. 2, D-37191 Katlenburg-Lindau, Germany
\and School of Space Research, Kyung Hee University, Yongin, Gyeonggi 446-701, Korea
\and Department of Physics, Swiss Federal Institute of Technology ETHZ, Zurich, Switzerland}

\date{}

\abstract {}
{Miyake et al. (henceforth M12) recently reported, based on $^{14}$C data, an extreme cosmic event in about AD775.
Using a simple model, M12 claimed that the event was too strong to be caused by a solar flare within the standard theory.
This implied a new paradigm of either an impossibly strong solar flare or a very strong cosmic ray event of unknown origin that
 occurred around AD775.
However, as we show, the strength of the event was significantly overestimated by M12.
Several subsequent works have attempted to find a possible exotic source for such an event, including a giant cometary impact
 upon the Sun or a gamma-ray burst, but they are all based on incorrect estimates by M12.
We revisit this event with analysis of new datasets and consistent theoretical modelling.}
{We verified the experimental result for the AD775 cosmic ray event using independent datasets including $^{10}$Be series
 and newly measured $^{14}$C annual data.
We surveyed available historical chronicles for astronomical observations for the period around the AD770s to identify
 potential sightings of aurorae borealis and supernovae.
We interpreted the $^{14}$C measurements using an appropriate carbon cycle model.}
{We show that: (1) The reality of the AD775 event is confirmed by new measurements of $^{14}$C in German oak;
(2) by using an inappropriate carbon cycle model, M12 strongly overestimated the event's strength;
(3) The revised magnitude of the event (the global $^{14}$C production $Q$=(1.1--1.5)$\cdot 10^8$ atoms/cm$^2$)
 is consistent with different independent datasets ($^{14}$C, $^{10}$Be, $^{36}$Cl) and can be associated with a strong,
 but not inexplicably strong, solar energetic particle event (or a sequence of events), and provides the first definite evidence
 for an event of this magnitude (the fluence $>30$ MeV was about $4.5\cdot 10^{10}$ cm$^{-2}$) in multiple datasets;
 (4) This interpretation is in agreement with increased auroral activity identified in historical chronicles.}
{The results point to the likely solar origin of the event, which is now identified as the greatest solar event
  on a multi-millennial time scale, placing a strong observational constraint on the theory of explosive energy
releases on the Sun and cool stars. }

\maketitle

\section{Introduction}

It is of particular interest, both from astrophysical and societal points of view, to understand the full spectrum of severity
 of extreme events in the Earth's radiation environment.
In particular, knowing the maximum possible energy of solar energetic particle (SEP) events and the
 frequency of their occurrence is of great importance for solar and stellar physics \citep{hudson10}.
The history of direct solar observations is relatively short, spanning only several decades and provides insufficient
 statistics on the occurrence rate of the most energetic SEP events \citep{smart06}.
Instead, long-term records of cosmogenic isotopes, like $^{14}$C and $^{10}$Be stored in terrestrial, meteoritic, or lunar archives, can reveal
 the history of such events over a greatly extended time-span \citep{usoskin_LR_08,beer12}.
While the statistics of extreme SEP events remains unclear to some extent \citep{hudson10}, progress in this field
 have been recently made \citep{schrijver12,usoskin_ApJ_12} allowing the occurrence frequency of such events to be
 better estimated by joint analysis of different cosmogenic isotope data.

\vspace{-0.2cm}
A particularly exciting result has been recently published by \citet[][ henceforth M12]{miyake12} who found
 a significant enhancement of about 1.5\% (15 permill)
 of $^{14}$C content measured in Japanese cedars around AD775.
Using a basic four-box model of the carbon cycle, M12 estimated the corresponding absolute global $^{14}$C production for the event
 as $6\cdot 10^{8}$ atoms/cm$^2$ or $19$ atoms/cm$^2$/sec averaged over a year.
This is an order of magnitude greater than the average $^{14}$C production rate due to galactic cosmic rays (GCR),
 which is estimated to be 1.6--2 atoms/cm$^2$/sec \citep[][and references therein]{kovaltsov12} for the pre-industrial era.
When translating the production rate into the flux of cosmic rays or the energy of the source, candidates being
 either a giant solar eruption or a nearby supernova, M12 concluded that
 it was much too high, implying a sudden strong cosmic ray event of unknown origin.
In an attempt to resolve the situation \citet{allen12} proposed ``a supernova largely hidden behind a dust cloud ...
 The resulting supernova remnant would be invisible.''
However as discussed below, this interpretation is unlikely.
Recently \citet{melott12} recalculated the energy of a possible coronal mass ejection (CME)
 related to such a burst of SEPs and found a lower value.
But it is still too energetic to correspond to a realistic solar eruptive event, leaving open the problem of the
 possible source of the AD775 event.
\citet{eichler12} tried to link the event source to an impact of a cometary body upon the Sun producing
 a shock in the corona to sufficiently accelerate energetic particles.
\citet{hambaryan13} proposed a gamma-ray burst as a possible source.
Thus, a number of exotic sources have been proposed, because the most plausible one, the SEP event,
 looks un\-realistically energetic given the strength of the event as estimated by M12.
Instead, the M12 result disagrees with data on another cosmogenic radionuclide, $^{10}$Be in the Dome Fuji ice core \citep{horiuchi08}.
The two cosmogenic isotopes ($^{14}$C and $^{10}$Be) are formed as sub-products of the same process of a
 nucleonic-electromagnetic-muon cascade caused by energetic cosmic rays or $\gamma-$quanta in the Earth's atmosphere
 \citep[e.g.,][]{beer12}.
Thus, it is hardly possible to obtain a tenfold increase in the annual $^{14}$C production
 without a corresponding increase in $^{10}$Be data \citep{usoskin_GRL_SCR06}.
This indicates an inconsistency in the scenario proposed by M12.
This event was also analyzed by \citet{usoskin_ApJ_12} who found the strength of the event
  much lower than did M12.
Here we revisit the AD775 event with analysis of new datasets and consistent theoretical modelling.

\section{$^{14}$C in German oak}

\begin{figure}
\centering \resizebox{7cm}{!}{\includegraphics{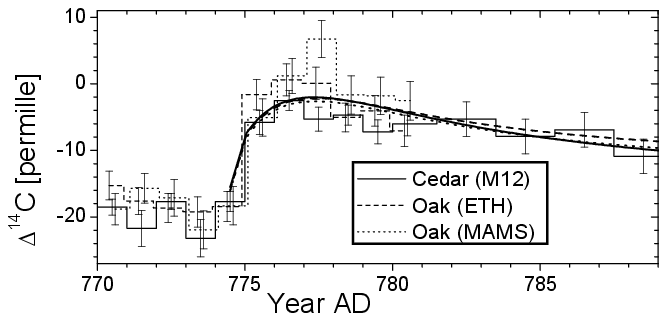}}
\caption{Time profiles of the measured $\Delta^{14}$C content in Japanese cedar \citep{miyake12}
 and German oak (this work) for the period around AD775.
 Smooth black and grey lines depict a family of best fit $\Delta^{14}$C profiles, calculated
  using a family of realistic carbon cycle models for an instantaneous injection of $^{14}$C into the stratosphere. }
\label{Fig:14C}
\end{figure}

In order to verify the existence and strength of the claimed AD775 event, annual samples from
 a German oak (tree Steinbach 91 from the river Main), a
 part of the German Oak Chronology \citep{friedrich04}, were measured independently in facilities of
 Mannheim, Germany (MAMS) and the ETH Zurich, Switzerland (ETH) for the period AD770--780.
The data (Fig.~\ref{Fig:14C}) are presented here for the first time.
One can see that they confirm the increase of $^{14}$C production around AD775 found by M12.
Despite a possible slight vertical offset explainable by local peculiarities,
 the magnitude and timing of the increase are in full agreement.
This implies that the event was global and was caused by the enhanced production of $^{14}$C.

\section{Analysis of $^{14}$C data}

\begin{figure}
\centering \resizebox{7cm}{!}{\includegraphics{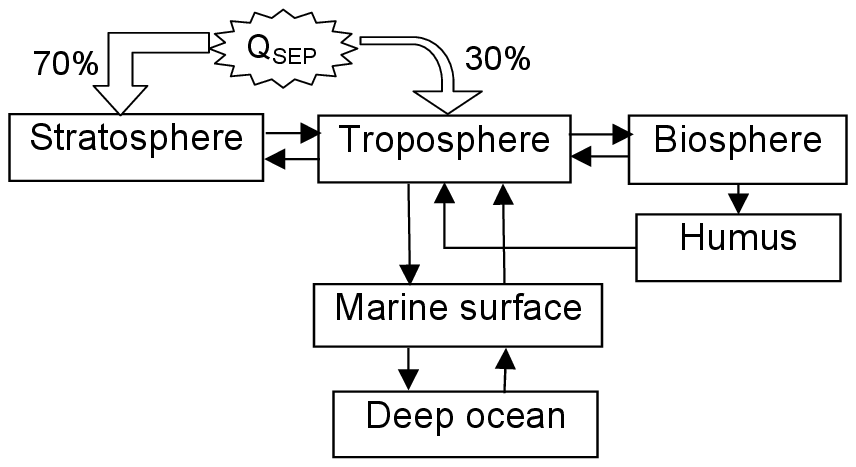}}
\caption{A scheme of the carbon cycle models used here with arrows depicting the carbon
 exchange between different reservoirs.
Big arrows denote production $Q$ of $^{14}$C in the atmosphere.}
\label{Fig:scheme}
\end{figure}

Only the relative content of $^{14}$C is measured in the atmosphere.
In order to evaluate the $^{14}$C production, and hence the strength of the event, one needs to
 run a carbon cycle model and to compare the relative increase of $\Delta^{14}$C with the background
 concentration of $^{14}$C in the troposphere, from where it is absorbed by living trees.
Here we use a family of three carbon cycle models, including the main carbon reservoirs (Fig.~\ref{Fig:scheme}):
 (1) a five-box model \citep[][page 703]{dorman04}; (2) a five-box model \citep{dergachev91,damon04},
 and (3) a box-diffusion model \citep{oeschger74,siegenthaler80}.
For better modeling of the SEP signal we also divided the atmosphere into stratosphere and troposphere,
 with a stratospheric carbon capacity of 15\% \citep[cf.][]{miyake12} and a residence time of two years \citep{damon78}.
A model \citep{kovaltsov12} was used to simulate the $^{14}$C production by SEPs with a hard energy spectrum,
 assuming 70\% of $^{14}$C produced by SEPs in the stratosphere and 30\% in the troposphere, globally.
These carbon cycle models differ slightly in exchange times between reservoirs and in dealing with the ocean,
 but they yield very similar results for the expected $^{14}$C signal (Fig.~\ref{Fig:14C}).
The background $^{14}$C level was taken as the mean pre-industrial production
 rate of 1.6 atoms/cm$^2$/sec \citep{goslar01}.
The best fit by the weighted least-square method
 between the three datasets and the three models yields a net $^{14}$C production of $(1.3\pm 0.2)\cdot 10^8$ atoms/cm$^2$.
This is a factor of about 5 smaller than the M12 model ($6\cdot 10^8$ atoms/cm$^2$).
M12 used a different carbon cycle model, neglecting the deep ocean (Fig.~\ref{Fig:scheme}), which is the greatest reservoir
 containing 92--95\% of all carbon.
Thus, M12 greatly overestimated the background $^{14}$C concentration in the troposphere.
This led to an error when translating the relative increase of $\Delta^{14}$C into the $^{14}$C production and so
 the event strength.
We conclude that M12 overestimated the production by a factor of 4--6 because of the inappropriate model.
Accordingly, all energy/particle flux estimates based on the numbers given by M12 are incorrect by the same factor.

We propose that the total $^{14}$C production for the AD775 event corresponds to a SEP fluence ($>30$ MeV)
 of $F_{30}\approx 4.5\cdot 10^{10}$ cm$^{-2}$, with a hard spectrum as per the SEP event of 23 Feb. 1956 \citep[cf.][]{usoskin_ApJ_12}.
Since $^{14}$C is produced by more energetic particles, the value of $F_{30}$ depends on the assumed spectrum.
For example, by assuming a very soft SEP spectrum as per the event of Aug. 1972, one would obtain the $F_{30}$ fluence $1.8\cdot 10^{11}$  cm$^{-2}$.
More robust is the fluence of SEP with energy $>200$ MeV, which is $(5.5$--$11)\cdot 10^9$ cm$^{-2}$ for our scenario
 irrespective of the assumed SEP energy spectrum.

In addition to the annual $^{14}$C measurements in individual trees, discussed above, we use a five-year
 averaged INTCAL09 global $^{14}$C series \citep{reimer_09} (Fig.~\ref{Fig:data}A).
The proposed scenario is consistent with the data ($\chi^2{\rm /DoF}=0.68$).
A time shift of a few years can be explained by the filtering of the raw data in the INTCAL09 series \citep{hogg09}.

\section{$^{10}$Be in ice cores}

\begin{figure}
\centering \resizebox{7cm}{!}{\includegraphics{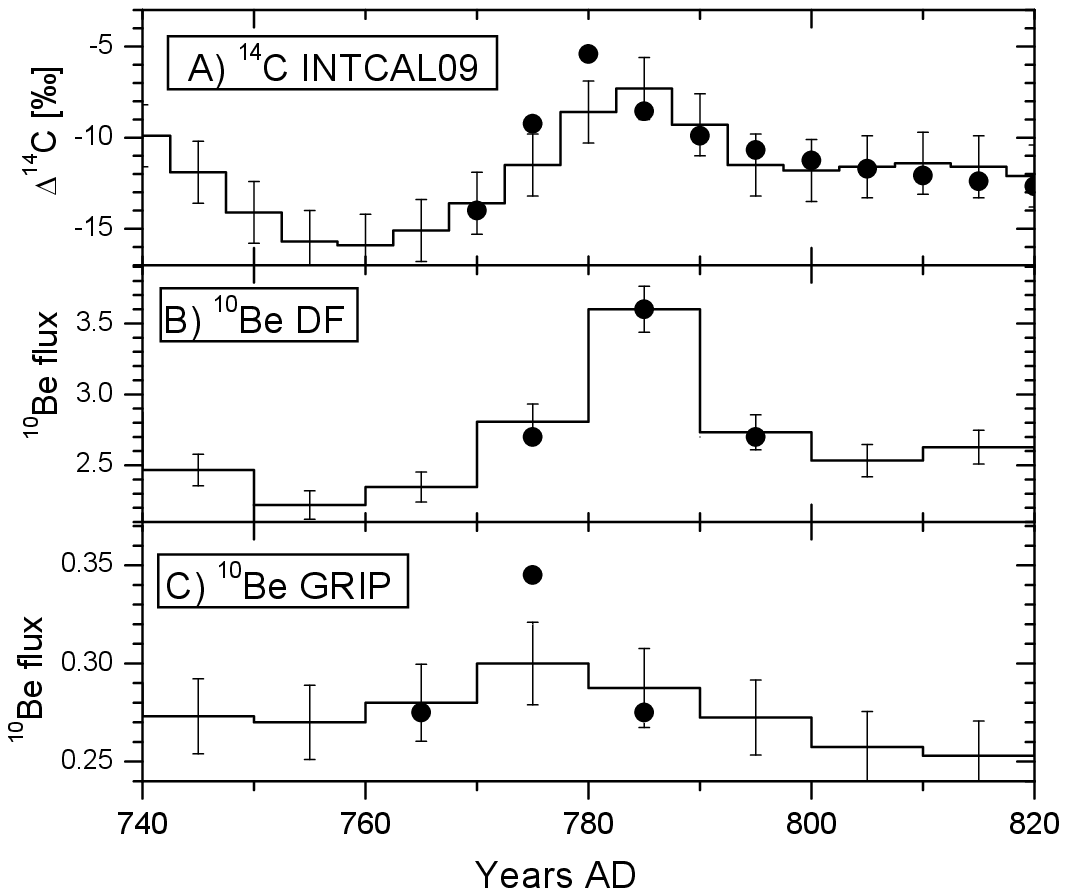}}
\caption{Time profiles of the measured cosmogenic isotopes (histograms with error bars) for the period around AD775.
Solid dots represent the enhancement expected in each data series assuming the AD775 event
 scenario discussed here ($^{14}$C production $Q=1.3\cdot 10^8$ at/cm$^2$).
A) INTCAL09 5-yr samples global atmospheric $\Delta^{14}$C \citep{reimer_09};
B) Quasi-decadal $^{10}$Be content in the Dome Fuji ice core \citep{horiuchi08}, our model data is
 shifted by 5 years to match the observed data;
C) Quasi-decadal $^{10}$Be content in the GRIP ice core \citep{yiou97}.}
\label{Fig:data}
\end{figure}

Using the above scenario, we calculated the expected increase in $^{10}$Be by applying
 the production model of \citet{kovaltsov_Be10_10} and assuming an instant injection
 and intermediate atmospheric mixing of $^{10}$Be \citep{mccracken_JGR_04,vonmoos06}.
Note that the ratio of the production of $^{14}$C and $^{10}$Be by a SEP event is almost independent
 of the assumption of the SEP spectrum \citep{usoskin_GRL_SCR06}.
Here we analyze two $^{10}$Be series measured in ice cores.
One is the Antarctic Dome Fuji series \citep{horiuchi07,horiuchi08} (Fig.~\ref{Fig:data}B).
The proposed scenario perfectly fits the data by magnitude, but
 the peak is delayed by several years and formally appears after AD780.
However, since the dating of the $^{10}$Be series is related to the resolution of the stratigraphic intervals sampled and
 is dependent on an age-depth model between tie points, dating uncertainties of several years
 are possible in the ice-core datasets \citep{beer00,horiuchi07}.
Thus, our proposed scenario is totally consistent with the $^{10}$Be data measured in
 the Antarctic Dome Fuji ice core.
The other dataset is the Greenland GRIP series \citep{yiou97,vonmoos06} (Fig.~\ref{Fig:data}C).
There is a weak increase during the AD770s but it is not pronounced.
The proposed scenario yields an expected peak that is higher (by about $2\sigma$) than the observed peak.
Thus, the GRIP series is not fully consistent with our scenario and the other data series,
 but the existence of the peak cannot be excluded at the 5\% level.
Instead, the corresponding $^{36}$Cl peak in the GRIP core\footnote{Estimate by JB based on unpublished preliminary data.} agrees with the proposed scenario.

\section{Historical observations of the Aurora Borealis}

We surveyed published aurora catalogues from oriental chronicles at low latitudes.
\citet{keimatsu73} and \citet{yau95} cite credible observations from Shanxi Province, China, in AD770 (twice), AD773, and AD775.
The next nearest observations are at AD767 and AD786.
We also survey catalogues from occidental chronicles.
\citet{link62} cites the ``red cross'' in the sky dated AD773/774 in different manuscripts of the Anglo-Saxon Chronicle (England),
 and ``inflamed shields'' in the sky (Germany, AD776).
The next nearest occidental observations are at AD765 and AD786.
At this time the Bible was a key reference in interpreting natural phenomena, explaining the cryptic reporting of aurorae
 (e.g., the above cited ``red cross'', also translated as ``red sign of Christ'' by Swanton 2000).
In a new survey of occidental chronicles, we identified probable aurorae in AD772 (``fire from heaven'', Ireland) and in an AD773
 apparition interpreted by Christians as riders on white horses (Germany).
Further new Irish observations are dated AD765 and AD786, and are found in the Annals of Ulster \citep{macairt83},
 and the above-cited AD773 report from Germany in the Royal Frankish Annals \citep{scholz70}.
We surveyed historical nova catalogues \citep[e.g.,][]{xu00} but found no credible report of supernovae in the AD770s.
\citet{allen12} interprets the red cross (Anglo-Saxon Chronicle) as an exotic nearby supernova with an unobservable
 remnant, but we interpret this as an aurora.
This is supported by a report that the same year ``snakes... seen extraordinarily
  in the land of the South-Saxons'' \citep{swanton00}.
Serpents often feature in descriptions of aurorae \citep{dallolmo80}, reflecting the sinuous movement of auroral structures.
We also note the uncertain dating of these events in the Anglo Saxon Chronicle, with \citet{swanton00}
 re-dating them to AD776, i.e., after the onset of the AD775 event.
We do not directly associate any particular aurora with the $^{14}$C event,
 but a distinct cluster of aurorae between AD770 and AD776
 suggests a high solar activity level around AD775.
With the next nearest observations around AD765-767 and AD786, this suggests an 11-year cyclicity.

\section{Discussion and Conclusions}

The existence of the AD775 event is confirmed using a larger dataset, including two new annual $^{14}$C series
 from German oak (Fig.~\ref{Fig:14C}), but the event's interpretation by M12 is found to be incorrect.
Because of an inappropriate carbon cycle model, M12 strongly overestimated (by a factor of 4--6) the strength of the event.
Here we re-conducted the analysis, using more realistic models.
The consequent event-integrated $^{14}$C production rate is (1.1--1.5)$\cdot 10^8$ atoms/cm$^2$.
We have verified that this value is in agreement with all the considered data series, including $^{10}$Be records in polar ice, yielding
 a consistent view corresponding to a strong SEP event (or a sequence of events) with a hard energy spectrum, 25--50 times stronger
 than the SPE event of 23 Feb. 1956 \citep[cf. ][]{usoskin_ApJ_12}.
Such an event, while very strong, is not impossible for the solar dynamo \citep{hudson10}.
Moreover, it can be a sequence of SEP events as, for example, happened in the Autumn of 1989, thus further reducing the
 severity of individual events.
This is corroborated by the steep tail of the SEP event fluence distribution \citep{schrijver12}.
Several potential candidates for similar events have previously been identified \citep{usoskin_ApJ_12},
 but observed in single data series only,
 thus leaving room for a possible terrestrial origin (e.g. regional climate excursion).
The AD775 event is the only one consistently observed in several independent datasets, thus providing the first
 unequivocal observational evidence of such a strong SEP event
 on a multi-millennial time scale.
This places a strong observational constraint on the upper limits of
 solar eruptive events, which is important for solar and more broadly stellar physics.

In conclusion, by correcting the M12 model, by providing new independent $^{14}$C data, and by surveying
 available historical chronicles and published aurora catalogues, we revisited the AD775 event
 to demonstrate that it can likely be attributed to a strong solar SEP event.
We show that:

* The existence of the AD775 event is confirmed by new measurements of $^{14}$C in German oak and by
  the existing $^{10}$Be data from polar ice cores.

* \citet{miyake12} overestimated the event's strength by a factor of 4--6.
 This directly affects subsequent works based on this incorrect estimate \citep[e.g.,][]{melott12,eichler12,hambaryan13}.

* The revised event is consistent with different independent datasets and is associated with a strong,
  but not inexplicably strong SEP event (or sequence of smaller events), providing the first definite evidence for a SEP
  event of this magnitude from multiple datasets.

* This interpretation is in agreement with enhanced auroral sightings reported in historical chronicles for the period.

\acknowledgement{
GAK was partly supported by the Program No.22 presidium RAS and by the Academy of Finland.}


\begin{thebibliography}{38}
\expandafter\ifx\csname natexlab\endcsname\relax\def\natexlab#1{#1}\fi

\bibitem[{{Allen}(2012)}]{allen12}
{Allen}, J. 2012, Nature, 486, 473

\bibitem[{Beer(2000)}]{beer00}
Beer, J. 2000, Space Sci. Rev., 94, 53

\bibitem[{Beer {et~al.}(2012)Beer, McCracken, \& von Steiger}]{beer12}
Beer, J., McCracken, K., \& von Steiger, R. 2012, Cosmogenic Radionuclides:
  Theory and Applications in the Terrestrial and Space Environments (Berlin:
  Springer)

\bibitem[{{Dall'Olmo}(1980)}]{dallolmo80}
{Dall'Olmo}, U. 1980, J. History Astron., 11, 10

\bibitem[{Damon {et~al.}(1978)Damon, Lerman, \& Long}]{damon78}
Damon, P., Lerman, J., \& Long, A. 1978, Ann. Rev. Earth Planet. Sci., 6, 457

\bibitem[{{Damon} \& {Peristykh}(2004)}]{damon04}
{Damon}, P.~E. \& {Peristykh}, A.~N. 2004, in AGU Geophysical Monograph Series,
  Vol. 141, Solar Variability and its Effects on Climate. Geophysical Monograph
  141, ed. {J.~M.~Pap et al.} (AGU, Washinton DC), 237--249

\bibitem[{Dergachev \& Veksler(1991)}]{dergachev91}
Dergachev, V. \& Veksler, V. 1991, Application of the Radiocarbon Method for
  Studies of the Environment in the Past (A.F. Ioffe Phys-Tech Inst., Acad.
  Sci. USSR, Leningrad, USSR (in Russian))

\bibitem[{Dorman(2004)}]{dorman04}
Dorman, L. 2004, Cosmic Rays in the Earth's Atmosphere and Underground
  (Dordrecht: Kluwer Academic Publishers)

\bibitem[{Eichler \& Mordecai(2012)}]{eichler12}
Eichler, D. \& Mordecai, D. 2012, Astrophys. J. Lett., 761, {L27}

\bibitem[{Friedrich {et~al.}({2004})Friedrich, Remmele, Kromer, Hofmann, Spurk,
  Kaiser, Orcel, \& Kuppers}]{friedrich04}
Friedrich, M., Remmele, S., Kromer, B., {et~al.} {2004}, {Radiocarbon}, {46},
  {1111}

\bibitem[{Goslar(2001)}]{goslar01}
Goslar, T. 2001, Radiocarbon, 43, 743

\bibitem[{Hambaryan \& Neuh\"auser(2013)}]{hambaryan13}
Hambaryan, V. \& Neuh\"auser, R. 2013, MNRAS, 429

\bibitem[{Hogg {et~al.}(2009)Hogg, Palmer, Boswijk, Reimer, \& Brown}]{hogg09}
Hogg, A., Palmer, J., Boswijk, G., Reimer, P., \& Brown, D. 2009, Radiocarbon,
  51, 1177

\bibitem[{Horiuchi {et~al.}(2007)Horiuchi, Ohta, Uchida, Matsuzaki, Shibata, \&
  Motoyama}]{horiuchi07}
Horiuchi, K., Ohta, A., Uchida, T., {et~al.} 2007, Nucl. Inst. Meth. Phys. Res.
  B, 259, 584

\bibitem[{Horiuchi {et~al.}(2008)Horiuchi, Uchida, Sakamoto, Ohta, Matsuzaki,
  Shibata, \& Motoyama}]{horiuchi08}
Horiuchi, K., Uchida, T., Sakamoto, Y., {et~al.} 2008, Quat. Geochronology, 3,
  253

\bibitem[{Hudson({2010})}]{hudson10}
Hudson, H.~S. {2010}, {Nature Phys.}, {6}, {637}

\bibitem[{Keimatsu(1973)}]{keimatsu73}
Keimatsu, M. 1973, A chronology of aurorae and sunspots observed in China,
  Korea and Japan, Part IV, Vol.~13 (Ann. Sci. College of Liberal Arts,
  Kanazawa University)

\bibitem[{{Kovaltsov} {et~al.}(2012){Kovaltsov}, Mishev, \&
  {Usoskin}}]{kovaltsov12}
{Kovaltsov}, G., Mishev, A., \& {Usoskin}, I. 2012, Earth Planet.Sci.Lett.,
  337, 114

\bibitem[{{Kovaltsov} \& {Usoskin}(2010)}]{kovaltsov_Be10_10}
{Kovaltsov}, G. \& {Usoskin}, I. 2010, Earth Planet.Sci.Lett., 291, 182

\bibitem[{Link(1962)}]{link62}
Link, F. 1962, Geofysik\'aln\'i Sborn\'ik, 173, 297

\bibitem[{Mac~Airt \& Mac~Niocaill(1983)}]{macairt83}
Mac~Airt, S. \& Mac~Niocaill, G., eds. 1983, The Annals of Ulster (to A.D.
  1131) (Dublin: Dublin Institute for Advanced Studies)

\bibitem[{McCracken(2004)}]{mccracken_JGR_04}
McCracken, K. 2004, J. Geophys. Res., 109

\bibitem[{{Melott} \& {Thomas}(2012)}]{melott12}
{Melott}, A.~L. \& {Thomas}, B.~C. 2012, Nature, 491, {E1}

\bibitem[{Miyake {et~al.}(2012)Miyake, Nagaya, Masuda, \& Nakamura}]{miyake12}
Miyake, F., Nagaya, K., Masuda, K., \& Nakamura, T. 2012, Nature, 486, 240

\bibitem[{Oeschger {et~al.}(1975)Oeschger, Siegenthaler, Schotterer, \&
  Gugelmann}]{oeschger74}
Oeschger, H., Siegenthaler, U., Schotterer, U., \& Gugelmann, A. 1975, Tellus,
  27, 168

\bibitem[{Reimer {et~al.}(2009)Reimer, Baillie, Bard, Bayliss, Beck, Blackwell,
  Ramsey, Buck, Burr, Edwards, Friedrich, Grootes, Guilderson, Hajdas, Heaton,
  Hogg, Hughen, Kaiser, Kromer, McCormac, Manning, Reimer, Richards, Southon,
  Talamo, Turney, van~der Plicht, \& Weyhenmeye}]{reimer_09}
Reimer, P.~J., Baillie, M. G.~L., Bard, E., {et~al.} 2009, Radiocarbon, 51,
  1111

\bibitem[{Scholz \& Rogers(1970)}]{scholz70}
Scholz, B.~W. \& Rogers, B., eds. 1970, Carolingian chronicles: Royal Frankish
  Annals and Nithard's Histories (Ann Arbor: University of Michigan Press)

\bibitem[{{Schrijver} {et~al.}(2012){Schrijver}, Beer, Baltensperger, Cliver,
  G\"udel, Hudson, McCracken, Osten, Peter, Soderblom, Usoskin, \&
  Wolff}]{schrijver12}
{Schrijver}, C.~J., Beer, J., Baltensperger, U., {et~al.} 2012, J. Geophys.
  Res., 117, {A08103}

\bibitem[{Siegenthaler {et~al.}(1980)Siegenthaler, Heimann, \&
  Oeschger}]{siegenthaler80}
Siegenthaler, U., Heimann, M., \& Oeschger, H. 1980, Radiocarbon, 22, 177

\bibitem[{{Smart} {et~al.}(2006){Smart}, {Shea}, {Spence}, \&
  {Kepko}}]{smart06}
{Smart}, D.~F., {Shea}, M.~A., {Spence}, H.~E., \& {Kepko}, L. 2006, Adv. Space
  Res., 37, 1734

\bibitem[{Swanton(2000)}]{swanton00}
Swanton, M., ed. 2000, The Anglo-Saxon Chronicles (London: Phoenix Press)

\bibitem[{Usoskin {et~al.}(2006)Usoskin, Solanki, Kovaltsov, Beer, \&
  Kromer}]{usoskin_GRL_SCR06}
Usoskin, I., Solanki, S., Kovaltsov, G., Beer, J., \& Kromer, B. 2006, Geophys.
  Res. Lett., 33, L08107

\bibitem[{Usoskin(2008)}]{usoskin_LR_08}
Usoskin, I.~G. 2008, Living Rev. Solar Phys., 5

\bibitem[{{Usoskin} \& {Kovaltsov}(2012)}]{usoskin_ApJ_12}
{Usoskin}, I.~G. \& {Kovaltsov}, G.~A. 2012, Astrophys. J., 757, 92

\bibitem[{Vonmoos {et~al.}(2006)Vonmoos, Beer, \& Muscheler}]{vonmoos06}
Vonmoos, M., Beer, J., \& Muscheler, R. 2006, J. Geophys. Res., 111, {A10105}

\bibitem[{Xu {et~al.}(2000)Xu, Pankenier, \& Jiang}]{xu00}
Xu, Z., Pankenier, D., \& Jiang, Y. 2000, East Asian archaeoastronomy:
  historical records of astronomical observations of China, Japan and Korea
  (Amsterdam: Gordon and Breach Science Publishers)

\bibitem[{Yau {et~al.}(1995)Yau, Stephenson, \& Willis}]{yau95}
Yau, K., Stephenson, F., \& Willis, D. 1995, A catalogue of auroral
  observations from China, Korea and Japan (193 B.C. - A.D. 1770), Technical
  Report RAL-TR-95-073, Chilton, Oxfordshire: Rutherford Appleton Laboratory

\bibitem[{Yiou {et~al.}(1997)Yiou, Raisbeck, Baumgartner, Beer, Hammer,
  Johnsen, Jouzel, Kubik, Lestringuez, Sti{\'e}venard, Suter, \& Yiou}]{yiou97}
Yiou, F., Raisbeck, G., Baumgartner, S., {et~al.} 1997, J. Geophys. Res., 102,
  26783

\end{thebibliography}

\end{document}